# Hall carrier density and magnetoresistance measurements in thin film vanadium dioxide across the metal-insulator transition

Dmitry Ruzmetov[1], Don Heiman[2], Bruce B. Claflin[3], Venkatesh Narayanamurti[1], Shriram Ramanathan[1]

[1]*Harvard School of Engineering and Applied Sciences, Harvard University, Cambridge, MA 02138, USA*

[2]*Department of Physics, Northeastern University, Boston, MA 02115, USA*

[3] *Semiconductor Research Center, Wright State University, Dayton, OH 45435, USA*

**Abstract**

Temperature dependent magneto-transport measurements in magnetic fields of up to 12 Tesla were performed on thin film vanadium dioxide ($VO_2$) across the metal-insulator transition (MIT). The Hall carrier density increases by 4 orders of magnitude at the MIT and accounts almost entirely for the resistance change. The Hall mobility varies little across the MIT and remains low, ~0.1cm$^2$/V sec. Electrons are found to be the major carriers on both sides of the MIT. Small positive magnetoresistance in the semiconducting phase is measured.

Vanadium dioxide is being actively investigated due to its potential in switching devices as well as fundamental scientific interest in understanding correlated electron systems. This compound undergoes a metal-insulator transition (MIT) upon temperature decrease through $T_{MIT}$=67˚C, as well as a sharp change in optical properties [1] and crystal lattice transformation near $T_{MIT}$. The importance of the contribution of electron correlations to the phase transition has been demonstrated [2], and mechanisms responsible for the MIT are being actively researched. In the Peierls model, the atomic distortion due to the lattice transformation at MIT from rutile metallic phase to monoclinic insulating phase causes the band gap opening [3]. In the Mott-Hubbard model, electron correlations alone can induce an insulator [4]. A correlation-assisted Peierls model where atomic structure aspects are considered on equal footing with intra-dimer V-V correlations has been suggested as well [5].

Recently, experiments by Cavalleri et al. based on ultrafast spectroscopy provided evidence for the bandlike character of the low-T insulator and suggested that the atomic arrangement of the high-T rutile unit cell is necessary for the formation of the metallic phase of $VO_2$ [6]. On the other hand, Kim et al. relied on femtosecond pump-probe measurements and temperature dependent XRD to put forward the picture where the metal-insulator transition and structural transformation from rutile to monoclinic lattice occur separately at different temperatures [7]. In this picture, there exists an intermediate metallic monoclinic phase between MIT and the structural phase transition. The fact that there is no lattice transformation to rutile phase at the MIT excludes the Peierls model and the driving mechanism of the MIT is considered to be the Mott transition. The origin of the metallic monoclinic phase was explained with hole-driven MIT theory [8, 9] and Hall effect measurements of the hole density were presented in support [7]. Evidence in favor of the Mott transition was presented also on the basis of infrared spectroscopy and nano-imaging [10].

Carrier density is an important parameter in the Mott theory and yet data for this parameter are very scarce, especially for thin films. Thin films are of particular interest for applications in electronics and electro-optic devices. Early Hall effect measurements in $VO_2$ showed that electrons were predominant carriers on the both sides of the MIT [11, 12, 13, 14, 15]. The values of the electron density reported by research groups could differ by 2 orders of magnitude both in the semiconducting phase at room temperature [11, 15, 13] and high-T metallic phase [11, 13]. As mentioned in the literature [15], the Hall effect measurements in $VO_2$ is a challenging task due to the following difficulties: low Hall mobility, high carrier density resulting in low Hall voltage, and unusually large amount of noise ascribed to be due to the strain present in the sample arising from the discontinuous lattice transformation at the structural phase transition (SPT).

When in the rare cases the temperature dependence of the carrier density across the MIT was measured in single crystal $VO_2$, low magnetic fields and non-direct methods (e.g. AC magnetic field) were used to overcome the noise issue mentioned above and low Hall signal, and the temperature dependence of the resistance displaying the MIT was not presented [15]. In light of this, direct DC measurements of the Hall effect in thin film $VO_2$ in high magnetic field accompanied with other electron transport data across the MIT are of great relevance. Carrier density determination in high-quality vanadium oxide films is particularly important, given the recent interest in exploiting the Mott transition for computational elements that may overcome limitations due to Si CMOS scaling [16].

In this paper we present the results of Hall and magnetoresistance measurements in thin film $VO_2$ in DC magnetic field of up to 12Tesla. The temperature dependence of the electrical resistivity, carrier density, and Hall mobility across the MIT are shown. The measured n-type conductivity in the semiconducting phase is shown to be consistent with recent photoemission spectroscopy results [1]. The novelty and importance of this work is that our high field measurement technique allowed for reliable Hall coefficient determination in both semiconducting and metallic states of high quality thin film $VO_2$.

Vanadium dioxide thin (~100nm thick) films were reactively DC sputtered in Ar (91.2%) + $O_2$ (8.8%) environment at 10mTorr from a V target. The base pressure in the sputtering chamber was $2x10^{-8}$ Torr. The substrate was kept at 550ºC during the deposition. Thin $VO_2$ films synthesized by this sputtering technique were comprehensively characterized by a variety of methods [1, 17] and the relationships between $VO_2$ film morphology and electron transport and band structure parameters were analyzed in our previous reports [18, 19]. In this work, electron transport measurements were performed on $VO_2$ films on c-plane $Al_2O_3$ substrates (samples A and B). $VO_2$ film on a sapphire substrate (sample A) was photo-lithographically patterned into a clover-leaf shape for van der Pauw measurements (Fig. 2a). The sample was silver paste mounted on a copper block equipped with a resistive heater. This custom made setup with LakeShore 340 temperature controller insured sample temperature stability within 0.05°C, which proved to be a necessary requirement since the sample resistance exponentially varied with temperature. Gold wires were indium-soldered to the sample. Another wiring method where a second layer of photo-lithography was used to connect 15μm-wide thin film gold leads to the $VO_2$ pattern yielded similar results. Electrical measurements on sample A were done with a DC current source and voltmeter. The copper block with the sample was placed inside a room temperature bore of a 14T cryogen-free magnet, Cryogenic Limited CFM-14T-50. Constant current $I_{24}$ was set through the sample A. The current magnitude was set to maximum before the current heating effects became present and varied from 35μA at 33˚C to 120μA at 64˚C in the semiconducting phase and up to 10mA in the metallic phase. Hall voltage $V_{13}$ was measured while the magnetic field was continuously ramped $0 \rightarrow 12\mathrm{T} \rightarrow -12\mathrm{T} \rightarrow 0$ at a preset temperature. The resistivity was measured by van der Pauw method and calculated by solving numerically the transcendental equation (1) in Ref.[20]. In order to determine the magnetoresistance, the voltage $V_{13}$ (in notations of Fig 2a) was measured at a constant current $I_{13}$ while the magnetic field was swept in positive and negative directions up to 12T in magnitude. The $V_{13}$ vs. B curves were fitted and averaged over all field sweeps (positive, negative sweeps and reverse directions).

X-ray diffraction (XRD) measurements were done on a Scintag 2000 diffractometer using Cu Kα radiation in θ-2θ geometry. The XRD spectrum from a thin vanadium dioxide film on Si substrate is displayed in Fig. 1. The d-spacing values of the observed peaks are inscribed in the figure and the $VO_2$ line assignment is done according to Israelsson *et al.* [21]. A comparison of the measured XRD spectrum with published data [21, 22, 18] indicates high-quality polycrystalline stoichiometric $VO_2$ with no detectable impurities.

In the magneto-transport measurements on the 12T apparatus, the temperature was incrementally increased through the MIT. At each constant temperature, the resistivity and the Hall voltage as a function of the magnetic field were measured. An example of the Hall voltage curves in semiconducting and metallic phases is shown in Fig. 2b. We present the results in terms of the Hall carrier density and mobility (Fig. 3) assuming a single band model as a first approximation. The values of the original measured Hall coefficient can be reproduced from the carrier density n(T) plot in Fig. 3 using the equation: $R_H = -1/(n\ e)$. Then the slopes of the $V_H(B)$ curves were used to extract Hall carrier densities using the equation (SI units): $n = I\ B\ /\ (V_H\ e\ d)$, where $I=I_{24}$ is the current through the sample, B – magnetic flux density directed perpendicular to the sample plane, $V_H=V_{13}$ – Hall voltage, $e=1.60\times10^{-19}$ C, $d=10^{-7}$ m – film thickness. The measured resistivity and carrier density for sample A are presented in Fig. 3. The sign of the Hall voltage indicates that electrons are the major contributors to the transport in both semiconducting and metallic phases, in agreement with prior results on single crystals [15].

The resistivity experiences a drop of over 3 orders of magnitude at the transition temperature $T_{MIT}$=70˚C, which is characteristic to vanadium dioxide. Together with the results of the x-ray diffraction analysis this demonstrates the high quality and stoichiometry of our synthesized $VO_2$ films. The mobility $\mu=1/(e\ n\ \rho)$ was determined from known resistivity ρ and carrier density n and plotted in the bottom panel of Fig 3. Positive magnetoresistance, i.e. a resistance increase upon application of magnetic field, was measured in the semiconducting phase at room temperature. Specifically, the magnetoresistance is $\Delta R/R = (0.09 \pm 0.02)\%$ in the ±12T field at 26˚C, where $R=V_{13}/I_{13}$ (in notations of Fig. 2a).

One can see in Fig. 3 that the Hall electron density increases by 4 orders of magnitude from $1.1\text{x}10^{19}\text{cm}^{-3}$ at 64˚C to $1.7\text{x}10^{23}\text{cm}^{-3}$ at 75˚C upon the MIT. The increase of the number of carriers accounts almost entirely for the decrease of electrical resistance which is also manifested in the small change in the mobility μ. Assuming the density of the vanadium ions to be $3\text{x}10^{22}\text{cm}^{-3}$ [23], the measured Hall carrier density in the metallic phase of a thin film corresponds to 4.7 itinerant carriers per vanadium ion and is consistent with previous reports for bulk single crystal $VO_2$ [15]. The fact that in the Hall measurements there appear to be more than one itinerant carrier per V ion may be explained by the presence of two types of conduction, n- and p-type, with electrons being the majority carriers [23].

The carrier density was carefully measured upon temperature increase at the onset of MIT. The electron density starts increasing continuously toward the value in the metallic state. The Hall effect measurements done with 12T technique on two different samples yielded consistent results. The error bars in Fig. 3 come from the uncertainties of the best slope $V_{Hall}(B)$ fit.

For sample B, a $VO_2$ thin film on $Al_2O_3$ substrate, the carrier density was estimated using an alternate experimental setup. In this apparatus, the temperature dependent Hall effect was measured with a fixed magnetic field of 1.4Tesla. An unpatterned film on a square (1cm x 1cm) substrate was wired for van der Pauw measurements. The data were collected using both polarities of current and magnetic field and were averaged to compensate for electromagnetic effects [24]. The calculated carrier density is displayed in Fig. 4. The resistivity is shown in the inset and exhibits the MIT of same order as for sample A. The noise is larger than in the 12T results (Fig. 3) which highlights the advantages of high magnetic field and patterning technique described above. The temperature dependence of the carrier density in the semiconducting phase below the MIT agrees for 12T and 1.4T data. There is an exponential rise of n up to the threshold of the MIT near 60˚C then n starts rising faster and there is a jump at 67˚C. The results in Figs. 3 and 4 agree also in the metallic phase above 80˚C. However the jump of the carrier density is sharper in the previous measurement. Whereas in Fig. 4, the MIT jump is smaller, little over 2 orders of magnitude, and n continues to increase up to 80˚C. The analysis of the $V_{Hall}$ vs. B curves measured by 12T sweeps method shows that the curves exhibit extra noise and deviate from linear behavior in the region from 67 to 80˚C. For this reason the data in Fig. 4 are important to take into account in order to evaluate the behavior of the carrier density near the MIT point.

In summary, Hall effect measurements were performed across the metal-insulator transition in polycrystalline thin film vanadium dioxide in fields up to 12T. The electron density is found to change from ~$10^{19}$ to $10^{23}$ $cm^{-3}$ at the MIT (comparable to that of single crystals) which accounts almost entirely for the drop in the resistivity. The positive magnetoresistance at room temperature is measured to be 0.09% in a 12T field.

We acknowledge NSF supplement PHY-0601184 for financial support. DR and SR are thankful to Dr. Jagadeesh Moodera (MIT) for several helpful discussions on the Hall measurements.

**Figures**

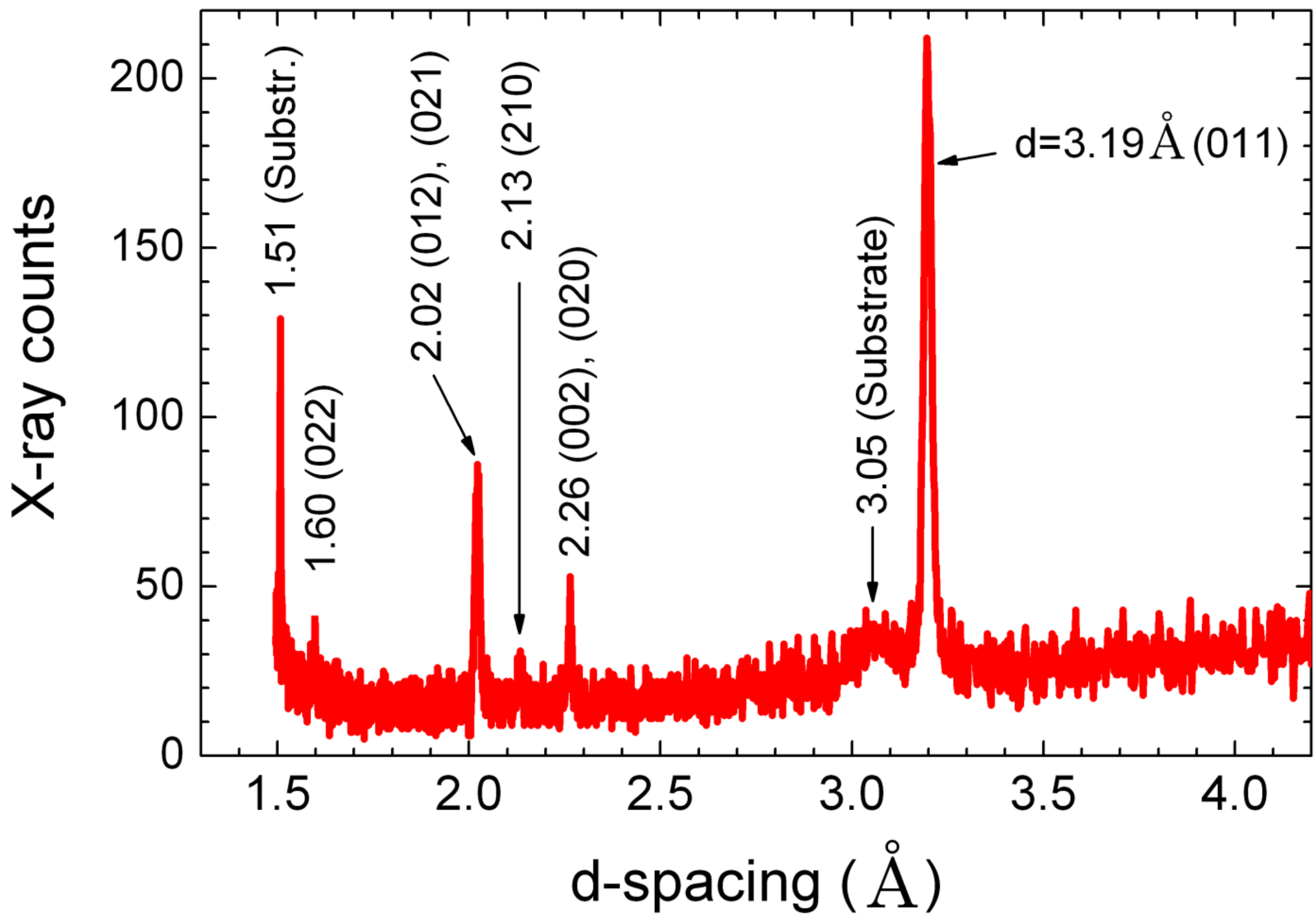

Figure 1. (Color online) XRD spectrum from a $VO_2$ thin film on Si(001)/$SiO_2$(native oxide) substrate. d values (in Å) of the peaks are inscribed and for $VO_2$ lines corresponding Miller indices of the Bragg planes are given in brackets according to Ref. [21].

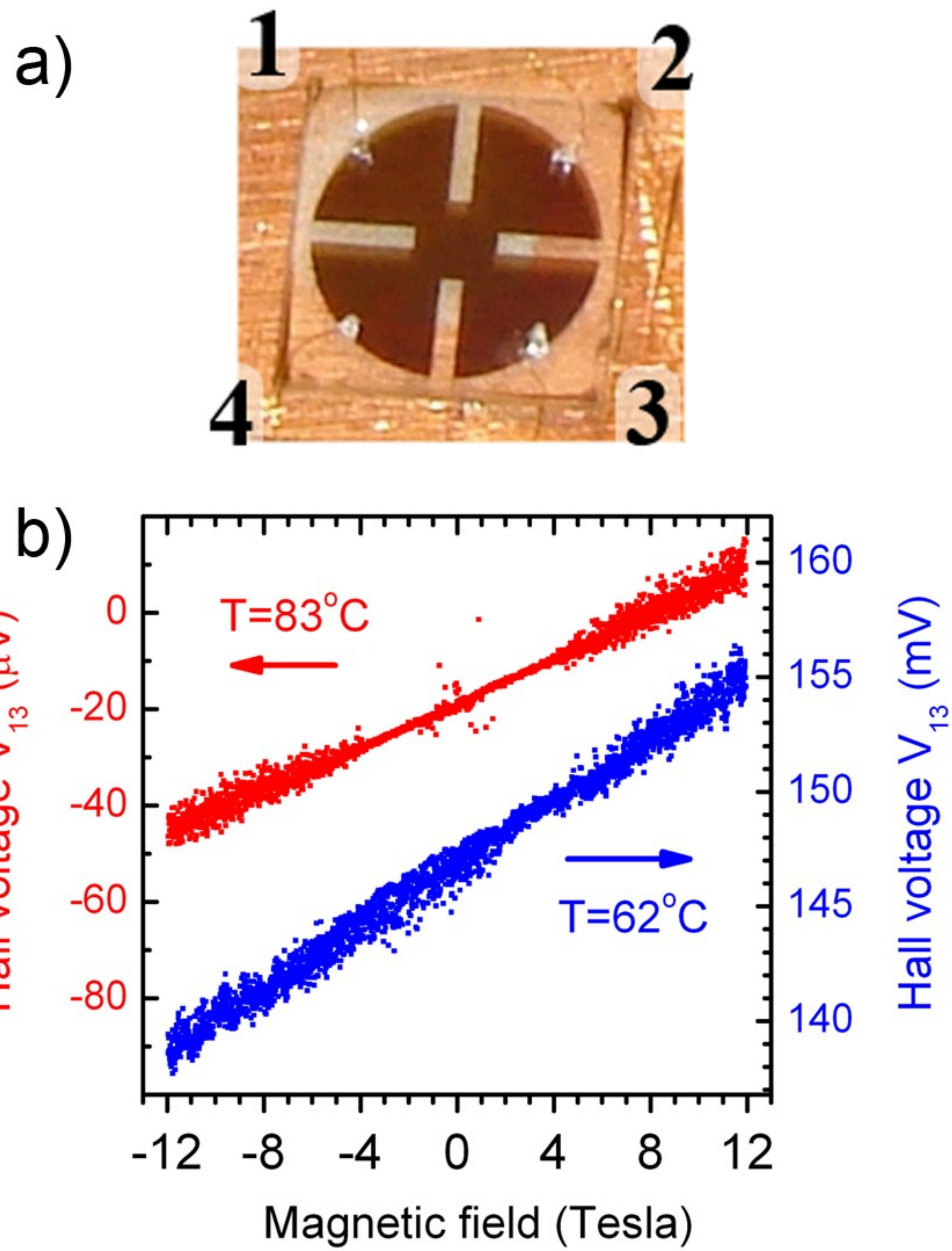

Figure 2. (Color online) **a)** $VO_2$ sample mounted on a copper base and set up for Hall measurements. Dark clover-leaf pattern is a $VO_2$ film on transparent square sapphire

substrate of 1cm x 1cm size. **b)** The Hall voltage $V_{13}$ in the metallic (upper red dot group) and semiconducting (lower blue dot group) state.

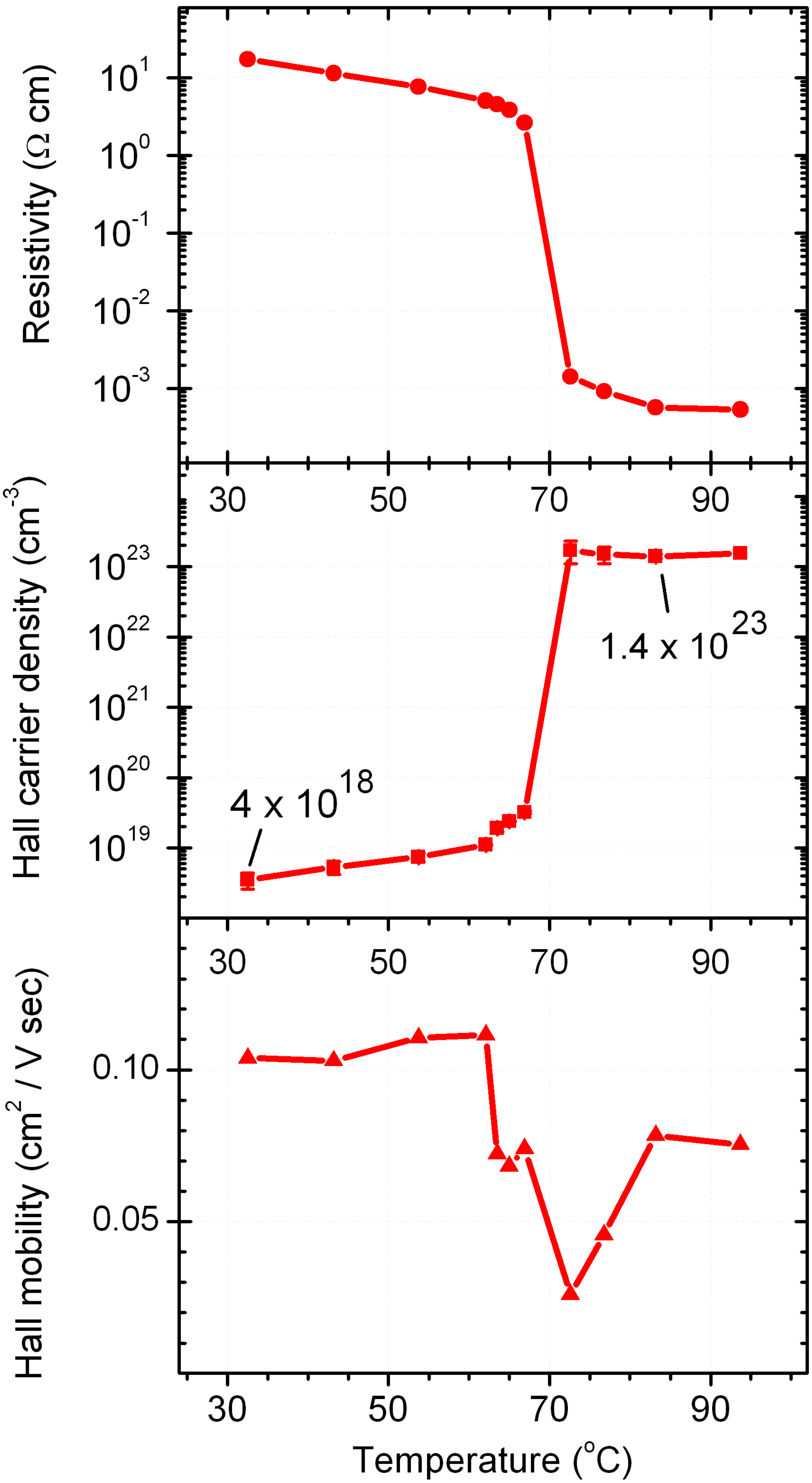

Resistivity (Ω cm)
10^1
10^0
10^-1
10^-2
10^-3
Hall carrier density (cm^-3)
10^23
10^22
10^21
10^20
10^19
1.4 x 10^23
4 x 10^18
Hall mobility (cm^2 / V sec)
0.10
0.05
30
50
70
90
Temperature (°C)

Figure 3. (Color online) Electron transport properties of a thin film $VO_2$ on an $Al_2O_3$ substrate (sample A) measured by 12T sweeping field apparatus. The Hall coefficient sign corresponds to electrons as the dominant current carriers.

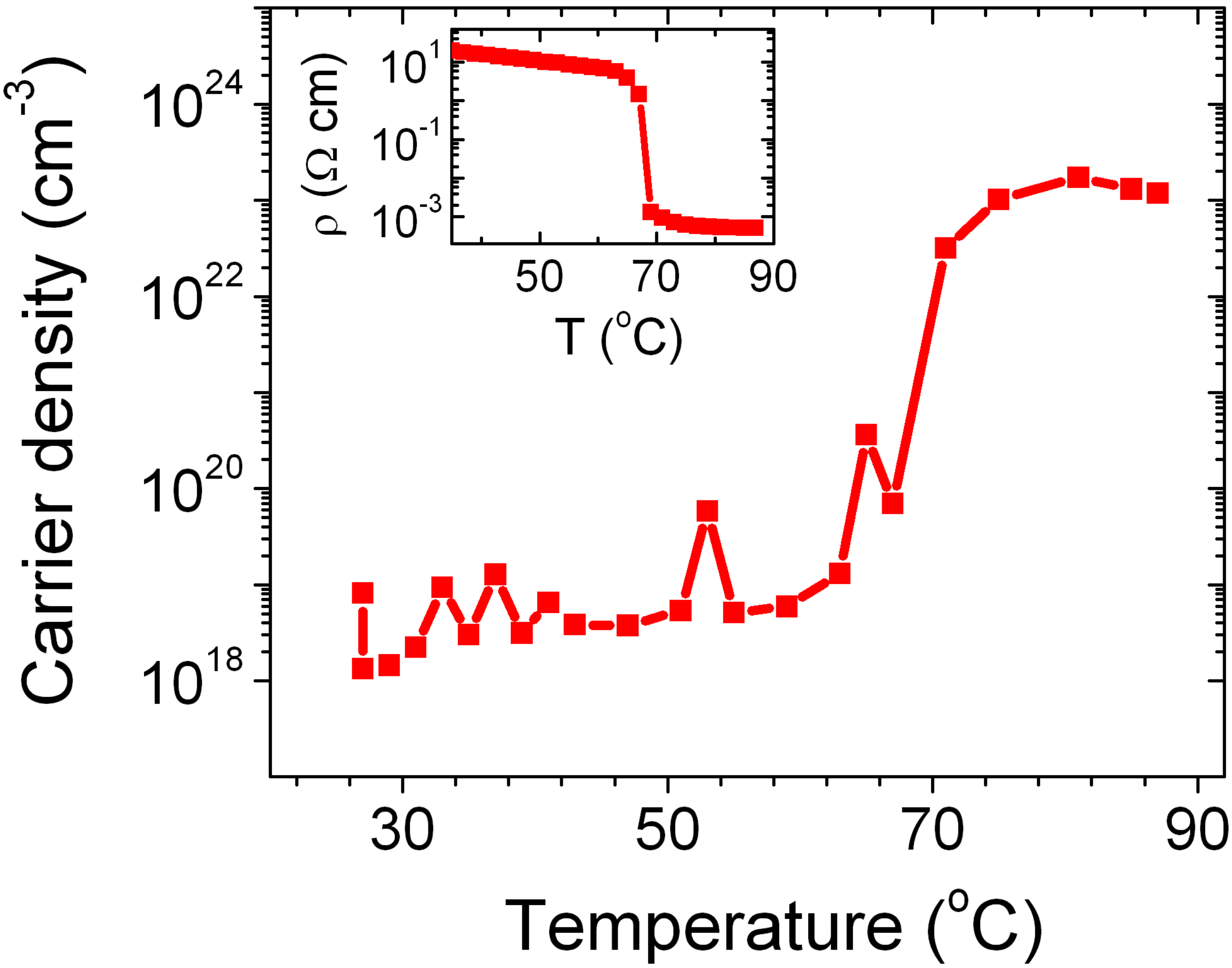

Figure 4. (Color online) Carrier density of a $VO_2$ film measured using 1.4T fixed field apparatus. The resistivity of the film is displayed in the inset.